# Robust and Interpretable Graph Neural Networks for Power Systems State Estimation

Arbel Yaniv
*Technical University of Munich*
Munich, Germany
arbel.yaniv@tum.de

Kilian Golinski
*Technical University of Munich*
Munich, Germany
kilian.golinski@tum.de

Christoph Goebel
*Technical University of Munich*
Munich, Germany
christoph.goebel@tum.de

*Abstract*—This study analyzes Graph Neural Networks (GNNs) for distribution system state estimation (DSSE) by employing an interpretable Graph Neural Additive Network (GNAN) and by utilizing an edge-conditioned message-passing mechanism. The architectures are benchmarked against the standard Graph Attention Network (GAT) architecture. Multiple SimBench grids with topology changes and various measurement penetration rates were used to evaluate performance. Empirically, GNAN trails GAT in accuracy but serves as a useful probe for graph learning when accompanied with the proposed edge attention mechanism. Together, they demonstrate that incorporating information from distant nodes could improve learning depending on the grid topology and available data. This study advances the state-of-the-art understanding of learning on graphs for the state estimation task and contributes toward reliable GNN-based DSSE prediction technologies.

## I. Introduction

State estimation (SE) is a core tool for grid operation and planning, aiming to recover the system state (bus voltage magnitudes and angles) governed by the AC power-flow equations, which are a set of $2(n-1)$ nonlinear equations that relate the system's bus voltages to the bus power injections. Unlike power flow analysis, which assumes perfect knowledge of loads and generation, SE aims to reconstruct the system state from imperfect measurements. This process is essential because comprehensive direct measurement at all of the grid's nodes is infeasible. For large-scale power grids, this problem consists of thousands of non-linear equations, thus computationally intensive to solve. Additionally, operational practices such as active control and frequent topology reconfiguration yields increased practical complexity of real-time operation, motivating the development of scalable and robust estimators. This motivates the use of machine learning models, which have been demonstrated to reduce computation time by one to two orders of magnitude [1].

Graph Neural Networks (GNNs) offer a natural framework for modeling power grids, which inherently exhibit a dynamic graph-structured nature. Their operational principle relies on message-passing mechanisms [2], [3], where node and edge embeddings represent system variables, and information is propagated across the network according to the grid connectivity [4]. Graph Attention Networks (GAT) enable neighborhood-specific weighting, attaining great success in many applications [5]. While GNNs show promise for DSSE, further investigation is needed for their reliable application for this task [?], [6], [7].

The ability to exchange information between nodes determines how effectively complex dependencies can be captured. Since each GNN layer propagates information through one-hop of neighbors, the straightforward approach to increase reach between nodes is to deepen the model architecture. However, increased depth often comes at the cost of degraded performance, a phenomenon traditionally attributed to oversquashing and over-smoothing [8]. Oversmoothing refers to the case where repeated message passing iterations make node embeddings similar, such as all buses in a power grid converging to similar voltage states. Oversquashing describes the loss of distant information when many signals are compressed through a finite-dimensional representation, e.g., when remote substations' influences are bottlenecked along with other information from nodes along the transmission path [8]. Recent theoretical advances reveal that these effects are symptoms of a more fundamental issue, the rank collapse, where node representations converge to a low dimensional subspace irrespective of the learned transformations [9], reducing the model's ability to represent diverse signals. The use of a Sum of Kronecker Products (SKP) is proposed [9] as a message passing scheme that prevents rank collapse and preserves multiple independent signals across layers, enabling richer representations of long-range interactions in graphs, thereby maintaining representation diversity and stable information propagation under challenging conditions such as sparse observations and varying graph structures. This property makes SKP a promising direction for improving GNNs predictions of DSSE, where accurate propagation of information over large spatial distances could be essential.

Potential modification to accommodate the meaningful exchange of information between nodes is especially important considering recent research on GNN-based state estimation suggesting that message passing beyond three to four hops does not contribute to learning the power-flow equations [10], leading to the conclusion that GNNs may not benefit from long-range propagation of information for this task. This motivates the analysis conducted in this study, aiming to examine whether this limitation arises from an inherent property of message passing learning capabilities of power-flow or rather from constraints within the employed architectures, and thus could be potentially mitigated. This question is natural considering that the power flow equations mathematically couple power injections at each bus with all other buses in the system through the admittance matrix. The present work investigates whether this limitation persists under variation to the message passing attention mechanism that is applied for the DSSE task. To facilitate the exploration of this question, the proposed architecture is examined via a by-design interpretable manner to enable explicit embedding contributions quantification. Beyond aiding the scientific analysis, interpretability is a key requirement for DSSE technologies where operators and regulators demand transparent decision logic [11]- [12]. This motivates a twofold research agenda: designing architectures that balance accuracy and interpretability, and developing diagnostics that explain why particular architectures perform better or worse on a given task.

## II. Related Work

A GNN architecture based on graph convolutional layers was proposed in [10] as a power-flow solver model. The evaluation included a sensitivity analysis with respect to the hop size, providing insights into the influence of information exchanges with both nearby and distant buses in the grid. The results indicated that message passing beyond three to four hops does not enhance learning performance for power grids, irrespective of their scale. This finding arises questions regarding the utilization of message passing for learning the power-flow equations, which will be examined in this study.

A purely data-driven GNN for power-flow prediction was presented in [13], indicating that the prediction error is lower for buses with high eigenvector centrality, empirically demonstrating that the proposed architecture's expressivity is linked to the spectral properties of the graph structure, that is, the eigenvalues and eigenvectors of the graph's matrix representation. The expressivity of a GNN refers to its ability to distinguish between different data structures and features. That is, expressive GNN should produce different representations for different inputs [14]. However, the simulations were conducted on constant topology only for the medium-scale NREL-118 power grid, limiting its through assessment.

A gated GNN was compared against dozen of GNN variants for learning the power-flow equations [15]. While showing promise compared to the other variants for small (30 bus) and large-scale systems (1354 bus), the model had high variance with high error for medium-scale grid (300 bus). This low reliability was prevalent across all GNN variants. Moreover, the analysis was limited only to single line outages, which does not offer a comprehensive experimental setup to truly examine the architecture's generalization. The limited exploration of the generalization performance is prevalent among other studies as well, which often does not include out-of-distribution experiments, where certain topological configurations are withheld during training and only used for testing [16]- [17]. While such works provide useful insights into the effect of architectural design choices on performance, the absence of validation under dynamic topologies neglects one of the key potential advantages of GNNs over MLPs: their potential to adapt to varying grid topologies, which is critical for real-world distribution grids.

While holding promise for modeling problems with dynamic graph structures, harnessing GNNs for learning the power-flow equations on changing grid topologies has been shown to have limited generalization beyond grid topologies seen during training [6], compromising their reliability for deployment within real-world operating technologies.

A graph convolutional model for DSSE was developed in [18], enhanced via feature scaling and pseudo-measurement generation process. The model achieves estimation accuracy comparable to computationally intensive sparsity-aware, model-based estimators while substantially reducing inference time. However, the experimental evaluation was restricted to fixed network topologies, with no switching events or topological variations considered, and the study did not include comparisons against standard multilayer perceptron (MLP) baselines, which constitute the primary leading alternative in fixed-topology learning settings. Similarly, a GNN architecture was benchmarked against a fully connected MLP architecture for the task of state estimation under varying grid configurations and noise contamination levels in [19]. The results show that GNNs outperformed the standard MLP only under high measurement penetration rates, while the MLP showed consistent reliability across variety of measurement penetration rates and topology changes scenarios, underscoring the need for further exploration of the GNN learning paradigm for the SE task.

Graph Neural Additive Networks (GNAN) [20] are motivated by the idea of combining the interpretability of generalized additive models [21] with the ability of GNNs to exploit structured dependencies. Instead of learning high-dimensional latent embeddings with limited transparency, GNAN decomposes predictions into a sum of feature-specific contributions that can be directly inspected. To account for the graph structure, these contributions are modulated by a distance-dependent weighting, allowing one to analyze how the influence of each node decays with graph distance. In order the pursue the examination of information exchange between nodes we employ GNAN in this study for the SE task for the first time. In this study, we evaluate predictive performance for voltage magnitude and phase angle while applying layer-wise diagnostics via GNAN to examine the representational changes that accompany information exchange depth. Moreover, we introduce a edge weighted Kronecker-based message-passing mechanism for GAT and GNAN (SKP-GAT and SKP-GNAN) as a controlled architectural intervention to examine potential performance gains. Overall, this study offers the following contributions:

- We introduce SKP-GAT and SKP-GNAN for DSSE, thereby enhancing the information exchange between nodes within the message passing framework.
- We utilize GNAN as an interpretable baseline for DSSE, demonstrating that predictions are predominantly influenced by immediate neighbors, but could benefit from distant nodes in certain cases.
- We compare the prediction accuracy of SKP-GAT and SKP-GNAN to that of basic GAT and MLP architectures to understand the impact of enhanced information exchange on model performance.

## III. Experimental Setup

Message passing in GNNs follows the principle that each node updates its hidden representation by aggregating transformed features from its neighbors. In this paper, we introduce several message passing variants to perform DSSE as follows.

### A. Graph Neural Additive Network

For interpretable graph leaning, we adopt the GNAN architecture [20]. Let $G = (V, E)$ with $|V| = N$ denote a graph, and $\mathbf{x}_i \in \mathbb{R}^d$ the feature vector of node $i$. For each feature dimension $k$, GNAN learns a univariate nonlinear function $f_k : \mathbb{R} \to \mathbb{R}$. In addition, a distance weighting function $\rho : [0,1] \to \mathbb{R}$ assigns importance to neighbors based on their shortest-path distance $\text{dist}(i, j)$. The representation of node $i$ in dimension $k$ is given by:

$$[h_i]_k = \sum_{j=1}^{N} \frac{1}{\#\text{dist}_i(j,i)} \cdot \rho\left(\frac{1}{1+\text{dist}(j,i)}\right) \cdot f_k([x_j]_k), \quad (1)$$

where $\#\text{dist}_i(j, i)$ can be understood as amount of nodes at the different distances $\text{dist}(j, i)$. This formulation ensures that all nodes at the same distance contribute equally, but their impact is rescaled by $\rho$.

For prediction tasks, GNAN aggregates these feature-wise representations. At the graph level, the prediction is computed with $\sigma(\cdot)$ denoting an activation function (identity for regression) and node embeddings are pooled as:

$$\mathbf{h} = \sum_{i=1}^{N} \mathbf{h}_i, \quad \hat{y} = \sigma\left(\sum_{k=1}^{d} [h]_k\right). \quad (2)$$

A key property of this setup is that predictions can be decomposed into interpretable contributions. The impact of node $j$ on feature $k$ can be explicitly quantified as

$$C_{j,k} = f_k([x_j]_k) \cdot \sum_{i=1}^{N} \frac{1}{\#\text{dist}_i(j,i)} \cdot \rho\left(\frac{1}{1+\text{dist}(j,i)}\right). \quad (3)$$

Equation (3) enables transparent attribution of both which features and which distances shape the model's prediction, which is particularly valuable in safety-critical domains such as power systems. Specifically, the aggregation of information is governed by a distance-based kernel $\rho(s)$, which assigns weights to neighboring nodes depending on their scaled hop distance. Close nodes have a stronger influence, while distant ones contribute less, making the spatial reach of information flow directly interpretable. This design explicitly links the message passing to the underlying grid topology, rather than learning it implicitly, following the concept of distance-weighted similarity in kernel methods [22]. In our experimental setup, we employ GNAN primarily as an interpretability probe, examining how learned distance weights $\rho$ govern the balance between local and long-range node influences.

### B. Graph Attention Network (GAT)

The Graph Attention Network (GAT) [5] computes attention coefficients that determine the relative importance of each neighbor during aggregation, rather relying on fixed topological weights. The layer-wise update is:

$$x_i^{(k+1)} = \sigma\left(\sum_{j \in \mathcal{N}(i)} \alpha_{ij}^{(k)} W^{(k)} x_j^{(k)}\right), \quad (4)$$

where $x_i^{(k)}$ denotes the feature vector of node $i$ at layer $k$ (the $i$-th row of $X^{(k)}$), $W^{(k)} \in \mathbb{R}^{d_k \times d_{k+1}}$ is a learnable weight matrix that transforms features to dimension $d_{k+1}$, $\mathcal{N}(i)$ is the set of neighbors of node $i$ (including $i$ itself), $\alpha_{ij}^{(k)}$ are the attention coefficients, and $\sigma$ is a nonlinearity function. The attention coefficients $\alpha_{ij}^{(k)}$ are normalized importance weights computed via a softmax over the neighborhood:

$$\alpha_{ij}^{(k)} = \frac{\exp(e_{ij}^{(k)})}{\sum_{\ell \in \mathcal{N}(i)} \exp(e_{i\ell}^{(k)})}, \quad (5)$$

where $e_{ij}^{(k)}$ are unnormalized attention scores. These scores measure the relevance of node $j$'s features to node $i$ and are computed using a shared attention mechanism:

$$e_{ij}^{(k)} = a^{(k)T} \text{LeakyReLU}\left(\mathbf{W^{(k)}} [x_i^{(k)} \| x_j^{(k)}]\right), \quad (6)$$

where $\mathbf{a}^{(k)} \in \mathbb{R}^{2d_{k+1}}$ is a learnable attention vector, and $\|$ denotes concatenation. The attention mechanism first transforms both node features through the shared weight matrix $W^{(k)}$, concatenates them, and applies a parametrized linear combination followed by LeakyReLU activation. This formulation corresponds to the GATv2 variant which is used in this study as the benchmark attention-based model [23].

### C. Message passing schemes: SKP, SKP-GNAN and SKP-GAT

*1) Summed Kronecker Product (SKP):* To enhance the expressivity and robustness of message passing in GNNs, we employ Summed Kronecker Product (SKP) principles to introduce the proposed variants of the examined native GNAN ans GAT models [9]. That is, distinct aggregation method can be employed for different features, and aggregation can be done independently from feature transformation. In the SKP framework, each layer computes:

$$X^{(k+1)} = \sum_{c=1}^{C} \tilde{A}_c^{(k)} X^{(k)} W_c^{(k)}, \quad (7)$$

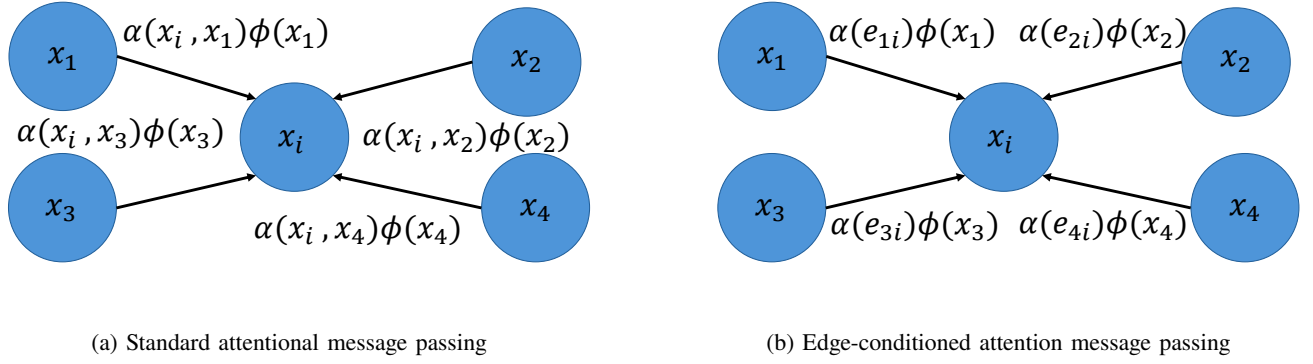

(a) Standard attentional message passing      (b) Edge-conditioned attention message passing

Fig. 1: Comparison of GNN architectures. (a) Attentional mechanism, where each incoming message is computed as $\phi(x_j)$ from the source node $j$ and scaled by an attention coefficient $\alpha(x_i, x_j)$ that depends on both the source node $j$ and the target node $i$. The aggregated neighborhood representation is given by $\bigoplus_{j \in \mathcal{N}_i} \alpha(x_i, x_j) \phi(x_j)$. (b) SKP attentional message passing mechanism, where each incoming message is computed as $\phi(x_j)$ from the source node $j$ and scaled by an attention coefficient $\alpha(e_{ij})$ that depends on the edge feature $e_{ij}$ connecting nodes $i$ and $j$. The aggregation is computed similarly to (a).

where $\tilde{A}_c^{(k)}$ are adjacency operators corresponding to each channel $c$, and $W_c^{(k)}$ are the respective feature transformation matrices. This can be equivalently written using the Kronecker product notation as:

$$\mathrm{vec}(X^{(k+1)}) = \sum_{c=1}^{C} \left( W_c^\top \otimes \tilde{A}^{(c)} \right) \mathrm{vec}(X^{(k)}), \quad (8)$$

where $\otimes$ denotes the Kronecker product:

$$(A \otimes B)_{(i,p),(j,q)} = A_{ij} B_{pq}. \quad (9)$$

The use of multiple channels enables the model to preserve multiple independent signal directions. Recent work demonstrates that edge-conditional transformations within the SKP framework enhance expressivity, improve robustness, and maintain diverse node representations [9], [24].

*2) SKP-GNAN: Edge-conditioned distance functions:* In the native GNAN architecture (Eq. 1), message passing is governed by a single distance-based weighting function $\rho(s)$, which defines a shared propagation operator across all features. This operator depends solely on hop distance and does not incorporate edge attributes. The proposed SKP-GNAN generalizes this formulation by introducing multiple channel-specific aggregation operators, while incorporating edge attributes. For each channel $c$, a learnable distance function $\rho^{(c)}$ is combined with edge-conditioned weights to construct a channel-specific adjacency operator $\tilde{\rho}^{(c)}$, defined as:

$$\tilde{\rho}_{ij}^{(c)} = \begin{cases} \rho^{(c)}(s_{ij}) \odot \mathrm{MLP}^{(c)}(E_{ij}) & \text{if } (i,j) \in E \\ \rho^{(c)}(s_{ij}) \odot \mathrm{MLP}^{(c)}(d_{ij}) & \text{otherwise} \end{cases} \quad (10)$$

where:
- $E_{ij}$ denotes the edge feature vector between directly connected nodes
- $\mathrm{MLP}^{(c)}(E_{ij})$ is a small MLP that computes edge-specific weights from edge features [24]
- $d_{ij}$ is the hop distance between non-adjacent nodes
- $\mathrm{MLP}^{(c)}(d_{ij})$ is a separate MLP that computes weights based on distance
- $\rho^{(c)}(s_{ij})$ is the distance function
- $\odot$ denotes elementwise product

For the distribution system state estimation task, we employ two SKP channels corresponding to voltage magnitude and phase angle.

*3) SKP-GAT: Edge-conditioned attention:* In the native GAT model the aggregation weights are computed from transformed node features, as shown in Eq. 6. Thus, the aggregation and feature transformation are coupled, and so the layer operator (mapping from GNN layer $k$ to the next layer $k+1$ as shown in Eq. 7) cannot be expressed as a sum of Kronecker products as shown in Eq. 8. In the proposed SKP-GAT, the attention mechanism is decoupled from node feature transformations by conditioning attention purely on edge attributes. Given node features $X^{(k)}$, edge attributes $E_{ij}$, and $H$ attention heads, the edge-conditioned attention weights are computed as:

$$z_{ij}^{(h)} = f_{\theta,h}(E_{ij}), \quad (11)$$

where $f_{\theta,h}$ is an MLP with learnable parameters $\theta_h$ that maps each edge feature vector $E_{ij}$ to a scalar weight for head $h$. This computation is independent of the node features and their transformations. The weights are then normalized via softmax:

$$[A_h(E)]_{ij} = \frac{\exp(z_{ij}^{(h)})}{\sum_{k \in \mathcal{N}(i)} \exp(z_{ik}^{(h)})}. \quad (12)$$

The resulting normalized matrices $A_h(E)$ are the aggregation operators in the proposed SKP-GAT formulation:

$$X^{(k+1)} = \phi\left( \sum_{h=1}^{H} A_h(E) X^{(k)} W_h \right), \quad (13)$$

where $\phi$ is a nonlinear activation function, and $W_h$ are head-specific feature transformations. We used $H = 3$ attention heads, corresponding to three SKP channels. As $A_h(E)$ are computed independently of $W_h$, the model can capture patterns that are invariant across different operational conditions. This design inherits the robustness and expressivity benefits of edge-conditional transformations [24], while the multi-head structure preserves multiple independent signal directions [9].

*D. Dirichlet energy and Rayleigh coefficient*

Understanding how layer-wise operations affect graph signals is crucial to understanding GNN learning. The *Dirichlet Energy* of a graph signal quantifies how smoothly node features vary across the graph structure, i.e., how similar connected nodes become during message passing [25], [26]. Given a graph $G = (V, E)$ with adjacency matrix $A$, degree matrix $D$, and the symmetrically normalized Laplacian $\Delta$, the Dirichlet Energy of node representations $X \in \mathbb{R}^{n \times d}$ is defined as

$$E(X) = \mathrm{tr}(X^\top \Delta X) = \frac{1}{2} \sum_{(i,j) \in E} \left\| \frac{x_i}{\sqrt{d_i}} - \frac{x_j}{\sqrt{d_j}} \right\|_2^2. \tag{14}$$

A low $E(X)$ indicates that neighboring node embeddings are similar, i.e., the representation is smooth with respect to the graph topology. As layers deepen in a GNN, an exponential decay of $E(X)$ could signal oversmoothing, where node states become nearly indistinguishable [8]. However, Dirichlet Energy alone cannot universally identify oversmoothing across architectures, as its behavior depends on model design and normalization choices. To disentangle magnitude effects from directional ones, the energy is normalized to yield the *Rayleigh Quotient* [27]:

$$\mathrm{RQ}(X) = \frac{\mathrm{tr}(X^\top \Delta X)}{\|X\|_F^2}. \tag{15}$$

This normalized measure captures the spectral properties of node embeddings: high values indicate rapid changes between connected nodes (high-frequency components), while low values reflect smooth, low-frequency behavior. The Rayleigh Quotient measures how much neighboring nodes differ relative to the overall size of the embeddings, revealing whether differences between nodes disappear because all features shrink uniformly or because all nodes actually become the same. For example, consider two connected nodes with embeddings $x_1 = 10$ and $x_2 = 12$. The Dirichlet Energy is proportional to $(10 - 12)^2 = 4$. Suppose that after several GNN layers, all embeddings are uniformly scaled by a factor of 0.1, yielding $x_1 = 1$ and $x_2 = 1.2$. The Dirichlet Energy drops to $(1 - 1.2)^2 = 0.04$, which could be misinterpreted as oversmoothing if Dirichlet Energy were considered alone. However, the ratio between the node difference and the overall embedding magnitude remains the same: $\frac{|10-12|}{\sqrt{10^2+12^2}} \approx \frac{|1-1.2|}{\sqrt{1^2+1.2^2}}$. The Rayleigh Quotient captures this normalized notion of difference and therefore remains approximately unchanged, indicating that the nodes have not become indistinguishable and that the decrease in Dirichlet Energy reflects uniform scaling rather than true oversmoothing. In our study, both $E(X^{(k)})$ and $\mathrm{RQ}(X^{(k)})$ are tracked across $k$ layers for GAT model variants.

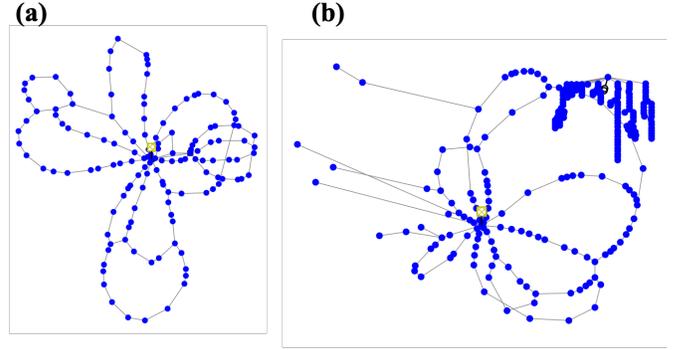

fig. 2: Visualized SimBench Grids. The two grids resemble a (a) 144 bus medium voltage grid and a coupled (b) 248 bus medium and low voltage grid

*E. Data Generation*

For our experiments we relied on the open-source SIM-BENCH dataset [28], which provides standardized distribution grid models. From this dataset we selected five representative grids of increasing size: `1-LV-rural-0-sw` with 15 buses, `1-LV-urban6-0-sw` with 59 buses, `1-MV-rural-1-sw` with 99 buses, `1-MV-urban-0-sw` with 144 buses, and `1-MVLV-semiurb-3.202-1-no_sw` with 248 buses. In Fig. 2, two of these networks are illustrated.

The simulated scenarios include the *switching* case from the DSML framework [19], in which redundant lines are alternated between on and off states to emulate grid reconfigurations due to outages or operational interventions. The number of switching events was randomly selected as a percentage of the grid's size as detailed in [19]. Specifically, the DSML methodology begins by randomly selecting a subset of buses in the grid where switching events will occur. The number of buses involved is determined probabilistically, drawn from a uniform distribution ranging from one bus up to twenty percent of the total buses in the grid. This ensures variability in the extent of topological changes while maintaining realism, since complete grid reconfiguration would be operationally impractical. For each selected bus, the framework evaluates whether redundant lines exist. A line is considered redundant if it can be switched off while still maintaining electrical supply to all buses in the network. This constraint is critical because it ensures that the switching events represent realistic operational scenarios rather than creating infeasible grid configurations that would result in power outages. The framework then alternates these redundant lines between active and inactive states to generate the switching events. The number of switching events per redundant line is also randomized, drawn uniformly between one and ten events throughout the simulation period (one year long data). In radial low-voltage networks, which lack connection redundancies [29], additional random lines were introduced to enable switching events, whereas medium-voltage networks already exhibit redundant connections [29] and therefore allow realistic switching operations. Learning alternating graph topologies is challenging, as models must generalize across multiple possible configurations, essentially performing out-of-distribution learning by encountering structural patterns not seen during training. Yet, this capability is a critical for modern distribution grid technologies, where topologies frequently change due to switching operations, faults, and regular maintenance.

To analyze the impact of grid observability, we considered measurement penetration rates of 0.2 and 0.9, corresponding to sparse and dense sensor coverage. We have also further conducted a sensitivity analysis for the stronger learning (attention-based) mechanism, covering the full observability spectrum in increments of 10%. The simulated measurement data used in this study based on the DSML framework [19] is modeled to represent two types of measurement devices with distinct precision specifications. Random noise following a normal distribution with zero mean and variance corresponding to error rate of 0.2% and 0.5% for voltage magnitude and angle measurements respectively, and 0.5%, 1.0% for active and reactive power measurements respectively is added to the state estimation parameters to represent Digital Local Network Stations (digiONS) measurement equipment for all transformers data in low-voltage grids and all measurements in medium-voltage grids. Similarly, according to the prevalent Intelligent Metering Systems (IMSys) devices that are deployed for household metering in Germany, voltage magnitude and angle errors of 0.5%, 1.0% respectively, and active, reactive power errors of 1.0%, 2.0% respectively is modeled for the low-voltage grids. The distinction between high-precision digiONS devices at critical infrastructure points and lower-precision household meters reflects actual deployment in distribution systems, making the results relevant for assessing method robustness under heterogeneous measurement quality conditions. The DSML framework utilizes load and generation profiles provided by [30] at a fifteen-minute resolution for an entire year, specifically using data from 2016. The resulting feature space, as fully described in [19] includes:

- **Node features**: including measurement features of voltage and power injections, parameter features relating voltage level and slack bus information and time related features.
- **Edge (line or transformer) features**: including power measurements along with their standard deviations, electrical parameters (i.e. conductance), operational parameters (i.e. phase shift) and switching status.
- **Labels**: bus voltages (magnitude and angle) obtained from power flow simulations, serving as ground truth for state estimation.

The GNAN model is built to operate with explicit structural information including: (1) a distance matrix and (2) a corresponding normalization matrix [20], which were calculated per grid topology as follows:

*Distance matrix.* For each simulated SimBench network [28], a NetworkX graph was built with all switches respected so that each switching scenario yielded the correct active topology. For each graph, we performed an all-pairs breadth-first search to obtain the unweighted shortest paths between any two buses. From these paths we derived the hop count $\ell_{ij}$, i.e., the number of edges between bus $i$ and bus $j$, and stored the result in a distance matrix. Following the scaling, the distances were then transformed to $s_{ij} = 1/(1+\ell_{ij})$, which maps adjacent nodes to $s = 0.5$ and decays smoothly towards zero for more distant nodes, while unreachable nodes remain zero [20].

*Normalization matrix.* A normalization matrix composed of elements $n_{ij}$ that counts, for each source bus $i$, how many target buses lie at the same hop distance $\ell_{ij}$ was calculated. This allows the models to normalize incoming messages per hop-shell, ensuring comparability across grids of different size and density. The resulting matrices $(s_{ij}, n_{ij})$ were stored as tensors and provided as `node_distances` and `norm_matrix` features to the GNAN variants during training [20].

### F. Model Training

All Datasets were constructed as PyTorch Geometric objects, following the standardized DSML preprocessing framework [31] and were split into 80% training, 10% validation, and 10% testing. Training was carried out using PyTorch Lightning with the Adamax optimizer, an initial learning rate of $10^{-3}$, and a learning-rate scheduler that reduced the step size upon stagnation of validation loss. An early stopping criterion was applied with a patience of three validation epochs to prevent overfitting. Performance was evaluated using the root mean square error (RMSE) between predicted and true bus voltage states, reported separately for voltage magnitudes and phase angles:

$$\text{RMSE} = \sqrt{\frac{1}{n}\sum_{i=1}^{n}(\hat{y}_i - y_i)^2}, \quad (16)$$

where $\hat{y}_i$ is the predicted value for sample $i$, $y_i$ is the true value, and $n$ is the number of samples. All models were tested across all five grids and both measurement rates (0.2 and 0.9), resulting in a diverse set of evaluation scenarios.

## IV. Experimental Results

### A. GNAN and SKP-GNAN

*1) Information exchange between nodes:* Comparing the native GNAN architecture with its edge-augmented variant, Fig. 3- 4 visualizes the learned distance functions $\rho^{(c)}(s)$ for the two largest simulated networks in our experiments (144 and 248 buses) for measurement penetration rate of 0.9 and 0.2, respectively. For the 144-bus grid, the native GNAN manages to learn a discernible gradient in its distance function, indicating gradually decreasing influence with hop distance. In contrast, for the 248-bus grid the function collapses to almost exclusively local interactions, with meaningful weights only for immediate neighbors. The SKP-GNAN variant, however, preserves the impact with a weak but noticeable gradient from near to distant buses in the 144-bus case and, for the 248-bus grid, maintains an approximately constant response up to the maximum hop distance of 49, indicating that long-range information is retained. As hypothesized, the edge-conditioned variant exhibits significantly higher activations at larger hop distances, showing that information from far away nodes is retained rather than being suppressed.

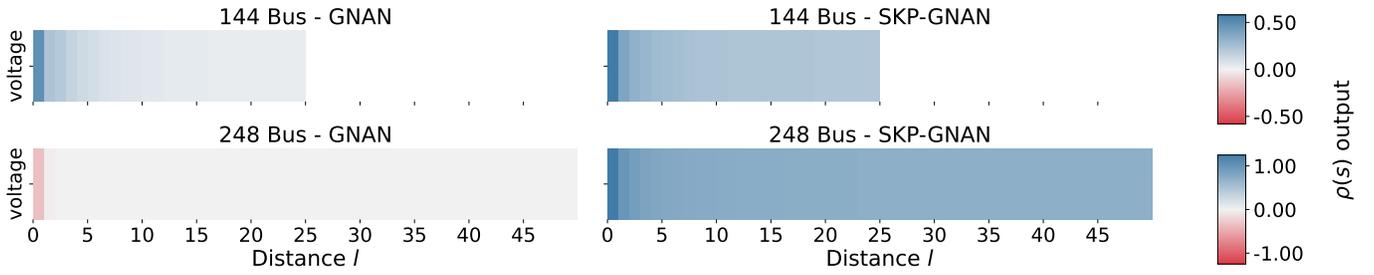

Fig. 3: Learned distance functions for a dense measurement penetration rate of 0.9 for the voltage magnitude task, evaluated on the two largest simulated networks (144 and 248 buses). For GNAN, the plotted curve corresponds to the learned distance function $\rho(s)$, for SKP-GNAN, $\tilde{\rho}^{(c)}(s)$. Higher absolute values indicate a stronger influence of features from nodes that are $\ell$ hops away.

*2) Predictive performance:* Concerning the assessment of the estimation accuracy, Fig. 5 shows boxplots of the resulting RMSE for voltage magnitude across all five examined grids and both measurement rates (0.2 and 0.9).

First, an analysis of the edge variant for the GNAN architecture is examined. As presented in Fig. 5, adding an edge conditioned MLP to GNAN slightly reduces the RMSE for high measurement penetration rate, attributed to the elimination of the high error outlier of the native GNAN model. The outliers in this figure correspond to the RMSE of the 248 bus grid, which shows significantly larger RMSE under all tested architectures. Considering the only near-range interactions for this grid as depicted in Fig. 3, it is evident that the restricted exchange of information between nodes has implications on the model's performance. Since the SKP-GNAN aggregates information from a wider set of nodes, its advantage grows when more measurements are available across the network. When grid observability is limited, exchanging information with other nodes is less beneficial, as with high probability, most nodes do not hold measured data. Therefore, aggregating information from more nodes is most advantageous under high measurement penetration rate. In contrast, under low measurement penetration, aggregating information from more distant nodes increases computational complexity without providing a proportional informational benefit. As only a small fraction of the nodes contribute meaningful measurement data, expanding the receptive field mainly incorporates uninformative signals.

### B. GAT and SKP-GAT

*1) Predictive performance under high measurement penetration rate:* Examination of the per-grid performance is shown in Fig. 6, which compares the RMSE of voltage magnitude for the GAT variants with the prevalent MLP benchmark [19].

At the high measurement rate (Fig. 6a), both GAT variants outperform the MLP baseline across all five examined grids, confirming that the attention-based architecture benefits from dense measurements. The SKP-augmented GAT matches the RMSE of the GAT in every grid, while showing improved performance for the 99 bus grid, which tends to higher RMSE over all models. This can be explained by the power-flow of this grid: its ground truth data has the highest amount of fluctuations in phase angle and voltage magnitude. This result highlights the robustness of the proposed SKP message passing mechanism for power-flow learning.

*2) Predictive performance under low measurement penetration rate:* At the low measurement rate (Fig. 6b), the picture changes. In this case the native GAT under-performs the MLP benchmark in all grids, reflecting its tendency to homogenize neighbor messages under sparse measurement conditions. In such sparse settings, attention weights are estimated from limited data, therefor losing its relative advantage over MLP. By contrast, the SKP-GAT variant manages to match the MLP benchmark in all networks but the largest, suggesting that the edge-weighted SKP mechanism successfully mitigates the performance degradation observed for the native GAT by reinforcing informative connections even when only few measurement nodes are available, as can be also seen in Fig. 5. This result underscores the benefit of the edge-weighted SKP mechanism within attention-based models under sparse measurement conditions, which is particularly important in real distribution systems which typically have limited observability. Similarly, for the larger grid (248 bus) as depicted in Fig. 8c, the MLP baseline performs best under low to medium measurement density, whereas for the high measurement density, the edge-weighted SKP variant performs slightly better compared to the other models. This can be also analyzed by the DE and RQ metrics in Fig. 7. For both GAT variants the Dirichlet energy remains at a similar value over depth, where the RQ for the SKP variant is higher compared to the native GAT, aligning with the slight observed performance gain.

For the phase-angle estimation, the results across all scenarios are presented in Table I. As can be seen, under high measurement penetration rate, the SKP-GAT and GAT exhibit comparable performance, with the SKP-GAT performing slightly better. Under sparse measurement, the MLP outperforms the attention-based models, consistent with the trends observed for the voltage magnitude estimation.

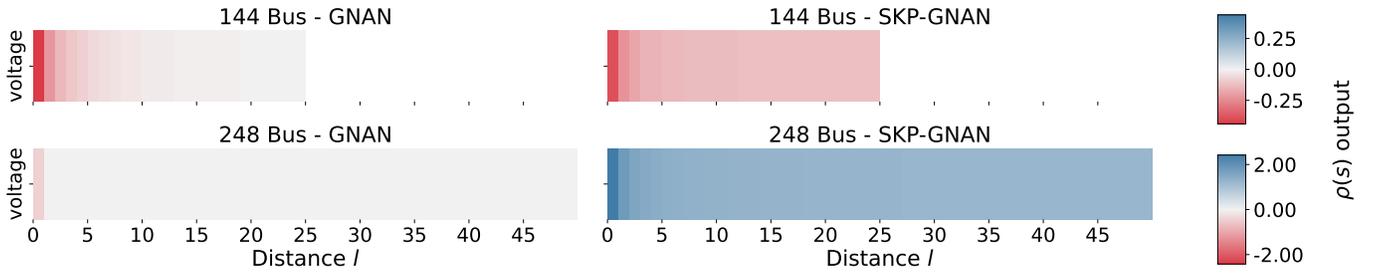

Fig. 4: Learned distance functions for a dense measurement penetration rate of 0.2 for the voltage magnitude task, evaluated on the two largest simulated networks (144 and 248 buses). For GNAN, the plotted curve corresponds to the learned distance function $\rho(s)$, for SKP-GNAN, $\tilde{\rho}^{(c)}(s)$. Higher absolute values indicate a stronger influence of features from nodes that are $\ell$ hops away.

TABLE I: RMSE results per grid and measurement rate (Magnitude / Angle).

| Grid | Nodes | Meas. | GNAN | | SKP-GNAN | | GAT | | SKP-GAT | | MLP | |
|---|---|---|---|---|---|---|---|---|---|---|---|---|
| | | | Magnitude | Angle | Magnitude | Angle | Magnitude | Angle | Magnitude | Angle | Magnitude | Angle |
| 1-MVLV-semiurb-3.202-1-no_sw | 248 | 0.2 | 0.006472 | 1.486 | 0.005254 | 15.49 | 0.00602 | 0.7777 | 0.005689 | 0.9304 | **0.003414** | **0.3376** |
| | | 0.9 | 0.00919 | 0.8554 | 0.007987 | 9.21 | 0.003214 | 0.5515 | **0.001956** | **0.3899** | 0.002609 | 0.5846 |
| 1-MV-urban–0-sw | 144 | 0.2 | 0.009687 | 5.703 | 0.009298 | 8.1 | 0.001672 | 0.2547 | 0.0008957 | 0.11 | **0.0007289** | **0.039** |
| | | 0.9 | 0.025 | 8.9 | 0.00775 | 17.27 | 0.000189 | **0.01891** | **0.0001457** | 0.02602 | 0.0005096 | 0.03223 |
| 1-MV-rural–1-sw | 99 | 0.2 | 0.00406 | 1.199 | 0.006297 | 0.817 | 0.003281 | 0.5375 | **0.001562** | 0.2717 | 0.00163 | **0.1521** |
| | | 0.9 | 0.007227 | 0.8157 | 0.003497 | 16.32 | 0.0003161 | 0.01964 | **0.0002425** | **0.01533** | 0.001556 | 0.1238 |
| 1-LV-urban6–0-sw | 59 | 0.2 | 0.001632 | 0.9018 | 0.002972 | 0.1523 | 0.000834 | 0.06929 | 0.0006118 | 0.06165 | **0.0002663** | **0.0208** |
| | | 0.9 | 0.001396 | 0.5138 | 0.003377 | 0.3169 | **0.0001215** | **0.01113** | 0.000138 | 0.01471 | 0.0003182 | 0.02203 |
| 1-LV-rural1–0-sw | 15 | 0.2 | 0.002502 | 1.44 | 0.004052 | 0.3625 | 0.0007953 | 0.0677 | **0.0006936** | 0.05658 | 0.0006949 | **0.04237** |
| | | 0.9 | 0.004477 | 1.64 | 0.005092 | 0.9795 | 9.964e-05 | 0.003745 | **9.855e-05** | **0.003727** | 0.0007054 | 0.03415 |

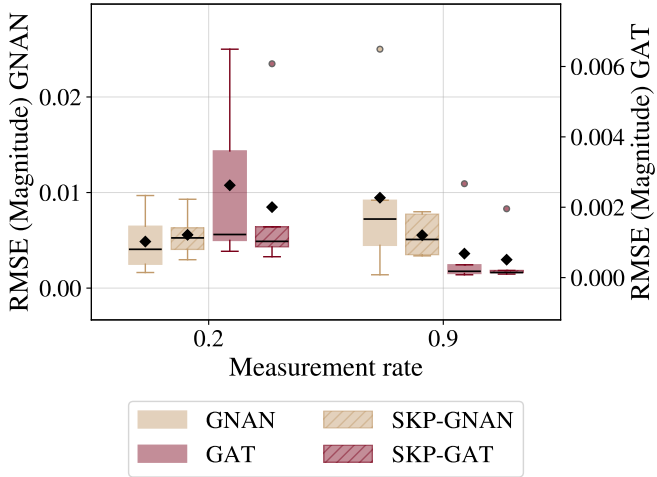

Fig. 5: Boxplots of RMSE (voltage magnitude) for GAT, GNAN and SKP-GAT, SKP-GNAN over all five grids

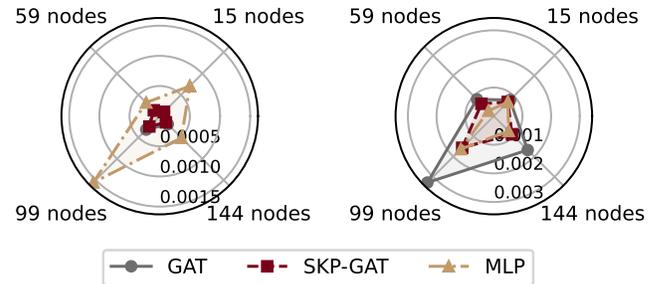

Fig. 6: Radar plots of RMSE (voltage magnitude) for GAT Native, SKP-GAT and MLP benchmark across the four smaller grids.

*3) Sensitivity analysis: varying measurement penetration rates:* A sensitivity analysis across varying levels of measurement penetration has been conducted, covering the full spectrum from 10% to full grid observability in increments of 10%. The results for the voltage magnitude are depicted in Fig. 8 for the 99-bus, 144-bus, and 248-bus grids, comparing

GAT, SKP-GAT, and the MLP baseline voltage magnitude performance. This analysis allows us to systematically evaluate model performance and robustness under different observability conditions and was also conducted for the voltage angle. In practical terms, varying measurement penetration rates can represent different levels of sensor deployment or data communication failures. The sensitivity study therefore provides insight into how each architecture behaves under partial observability and identifies the regimes in which the proposed SKP variant offers clear advantages over the native attention mechanism and the MLP baseline. Across all tested grids, the MLP performance remains relatively stable over the full range of measurement penetration rates. In contrast to the previously discussed comparison at two representative points (20% and 90%), the continuous analysis across the entire spectrum reveals clear and consistent trends: (1) attention-based mechanisms (both native GAT and SKP-GAT) become increasingly advantageous as observability increases, and (2) the benefit of the SKP mechanism is more pronounced under partial observability.

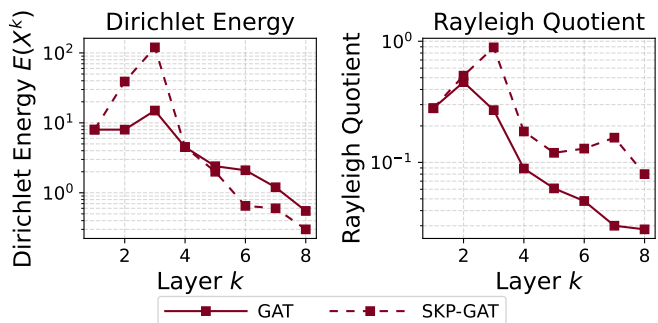

Fig. 7: Evolution of the Dirichlet Energy and Rayleigh Quotient across $k$ layers for GAT and SKP-GAT for the 248 bus grid under a measurement rate of 0.9

TABLE II: Training time (s), inference time (s), and number of parameters.

| Model | Meas. = 0.2 Train | Meas. = 0.2 Infer | Meas. = 0.9 Train | Meas. = 0.9 Infer | Params |
|---|---|---|---|---|---|
| GNAN | 4865 | 3.50 | 4858 | 4.28 | 1590 |
| SKP-GNAN | 21858 | 6.38 | 21683 | 7.67 | 3290 |
| GAT | 855 | 3.57 | 854 | 3.85 | 19106 |
| SKP-GAT | 679 | 3.11 | 866 | 3.19 | 5166 |
| MLP | **108** | **1.26** | **99** | **1.32** | 106000 |

Concerning the latter observed key trend, the results indicate that for the already strong native GAT mechanism, performance under high measurement penetration is primarily driven by sufficiently informative local measurements (as for high observability, with high probability local neighbors would have measurement information). Therefore, in this regime, the attention-based aggregation is already effective, and the additional edge-conditioned learning mechanism introduced by the proposed SKP leads to only marginal performance improvements over the native GAT. In contrast, for low observability, as the learning task becomes more challenging, the advantage of the SKP mechanism becomes more evident. Under partial observability, local information alone is insufficient, and the added learning mechanism of the proposed SKP-GAT becomes more beneficial. For the tested 99 and 144 bus grids, this performance gap is dominant up to 20-50% observability, and for the largest 248 bus grid, the SKP variant outperforms the native model over almost the entire spectrum except for the full observability scenario where their performance matches. Similar behavior was observed for the voltage angle estimation, as depicted in Fig. 9. Thus, for strong attention-based learning mechanisms, the relative gain introduced by SKP increases as measurement availability decreases. As across the full observability spectrum the proposed variant attain similar or better results compared to the native model for all tested grids, and considering that real distribution systems typically operate under partial observability rather than near-complete sensing, the SKP-augmented GAT provides a more robust and practically favorable estimator for DSSE compared to the native model, achieving comparable accuracy in dense scenarios and systematic gains when measurements are sparse.

Regarding the first mentioned identified trend, a key observation is that a distinct turning point can now be identified- the measurement penetration rate at which attention-based models consistently outperform the MLP baseline up to full observability. This transition occurs at different levels, directly proportional with the grid size: 60% for the 99-bus grid, 70% for the 144-bus grid and 80% for the 248-bus grid. These threshold values suggest that larger grids require higher measurement penetration before attention-based architectures can fully leverage their structural modeling capacity compared to MLPs. Below these thresholds, the MLP remains competitive due to its full observability, but with sufficient observability, the efficient attention mechanism consistently provides improved accuracy while being suitable in its design for graph-structure data, rendering the need for per-topology modeling of the learning architecture.

### C. Computational Efficiency

As shown in Table II, among the graph-based models, GAT and SKP-GAT exhibit considerably shorter inference times than the SKP-GNAN model, with the SKP-GAT being slightly faster. The modest reduction in training time of the SKP-GAT is consistent with its smaller parameter count. Although the MLP has the largest parameter count, it has the fastest training and inference as it is based on dense matrix multiplications, which are fully parallelizable.

## V. Discussion

The experimental results confirm our initial hypothesis that edge-conditioned message passing could contribute to the meaningful exchange of information between nodes within the receptive field. The heatmaps of the learned distance functions (Fig. 3) clearly show that SKP-GNAN preserves non-local activations even on the largest grids, while the native GNAN collapses to essentially local interactions. It can be also observed, as depicted in Fig. 5, that this enhanced neighbor exchange of information could translate into performance gains, depending on the available data, as demonstrated for the GNAN model under different measurement penetration rates.

Particularly, considering specific instances such as the voltage magnitude prediction via GNAN for the largest simulated grid (248 bus), this study provides empirical evidence that challenges previous findings in the literature regarding the role of multi-hop neighborhood information in GNN-based power flow prediction. While earlier work suggested that message passing beyond three to four hops does not contribute to learning power-flow, our analysis demonstrates that this limitation stems from constraints of the architecture rather than from fundamental properties of power-flow learning itself.

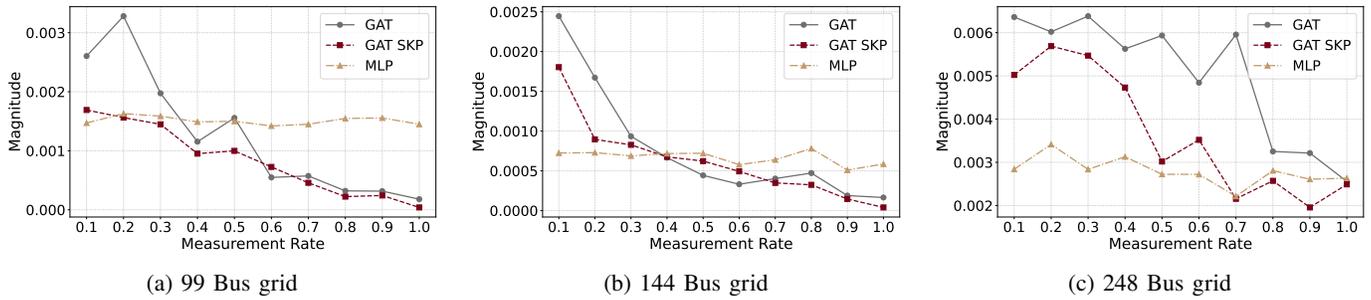

(a) 99 Bus grid     (b) 144 Bus grid     (c) 248 Bus grid

Fig. 8: Voltage magnitude RMSE across varying measurement penetration rates.

Through the application of an interpretable GNN architecture enhanced with the proposed edge-weighted implementation, we have shown that incorporating information from distant nodes across the network could yield measurable improvements in estimation accuracy, particularly in high measurement penetration scenarios. The results indicate that when the model architecture is designed to effectively aggregate information from a wider set of nodes, the receptive field can be successfully leveraged to improve learning outcomes. This finding aligns with the mathematical power-flow equations as power injections at each bus are coupled with all other buses in the system, confirming that architectural design choices rather than inherent structural limitations determine the effective range of information exchange in GNNs for the task of power-flow prediction.

## VI. Conclusions

This study demonstrates the effectiveness of different message passing mechanisms on the meaningful exchange of information between nodes. Through interpretable edge-attention SKP mechanism, we quantified the message passing framework's effect on estimation performance, which has been shown to depend on the underlying attention mechanism as well as the available information and the prediction task.

Moreover, the proposed edge conditional attention SKP has attained state-of-the-art performance in voltage magnitude prediction consistently for all simulated grids under high measurement penetration rate. These findings are aligned with recent theoretical and empirical work on edge-conditional message passing and support the use of the proposed edge-attention mechanism as a principled means to enhance GNN performance and robustness for DSSE.

While GNAN emphasizes transparency, its design inherently limits expressivity by restricting to univariate feature functions. Examination of extensions of GNAN that include the modeling of higher-order interactions and cross-feature dependencies would be a valuable direction to further explore for the continual development state estimation GNN-based models. Additionally, other message passing mechanisms with modifications to the attention and message calculations could be a promising research avenue, as demonstrated here for the suggested edge-conditioned attention. Furthermore, while this study simulates several low- and medium voltage distribution grids, including additional grid types and sizes is essential for demonstrating the scalability and reliability of GNNs as robust power-flow solvers.

All in all, our study provides the first systematic evaluation of edge-attention message-passing for DSSE. On the architectural side, we introduced a principled SKP augmentation for both tensor factorized (GNAN) and attention-based (GAT) estimators, implemented within the DSML framework. This study demonstrates that edge-conditional SKP message passing can be seamlessly integrated into existing graph architectures and yield tangible benefits. Beyond methodological advances, this work contributes to the existing literature by empirically demonstrating, through an interpretable edge-attention mechanism, that information exchange between distant nodes could enhance power-flow prediction performance. Thereby, the proposed study provides an important foundation for future research on GNNs for DSSE.

## VII. Acknowledgment

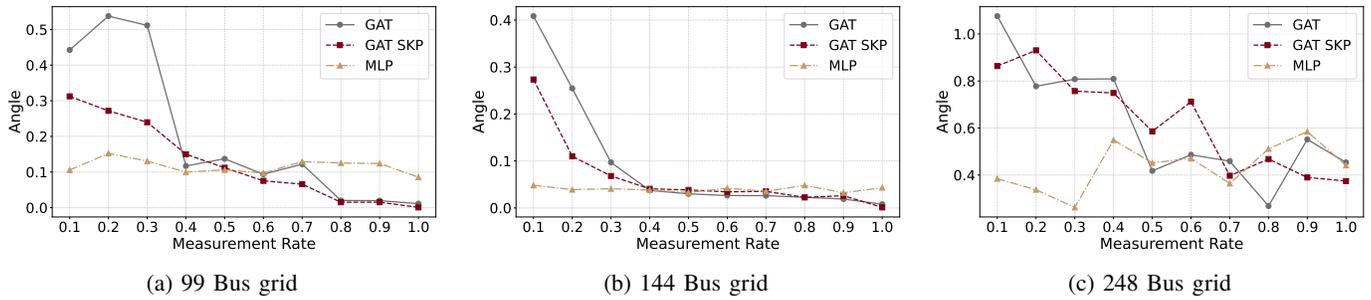

(a) 99 Bus grid     (b) 144 Bus grid     (c) 248 Bus grid

Fig. 9: Voltage angle RMSE across varying measurement penetration rates.